\RequirePackage{fix-cm}
\documentclass[twocolumn,epjc3]{svjour3}
\smartqed
\usepackage{cite}
\usepackage{isotope}
\usepackage[separate-uncertainty,retain-explicit-plus,per-mode=symbol,binary-units]{siunitx}
\DeclareSIUnit\cps{cps}
\DeclareSIUnit\electron{\mbox{$e^-$}}
\DeclareSIUnit\barg{barg}
\usepackage[colorlinks,citecolor=blue,urlcolor=blue]{hyperref}
\usepackage{lineno}

\usepackage[utf8]{inputenc}
\usepackage[T1]{fontenc} 
\usepackage{appendix}
\usepackage{todonotes}
\usepackage{amsmath}
\usepackage{wasysym}
\usepackage{nicefrac}
\usepackage{tabularx}

\DeclareMathSymbol{\mdot}{\mathord}{symbols}{"01}

\title{Direct comparison of PEN and TPB wavelength shifters in a liquid argon detector}
\newcommand{\AQLNGS}{INFN Laboratori Nazionali del Gran Sasso, Assergi (AQ) 67100, Italy}
\newcommand{\AQGSSI}{Gran Sasso Science Institute, L'Aquila 67100, Italy}
\newcommand{\AstroCeNT}{AstroCeNT, Nicolaus Copernicus Astronomical Center of the Polish Academy of Sciences, 00-614 Warsaw, Poland}
\newcommand{\BOINFN}{INFN Bologna, Bologna 40126, Italy}
\newcommand{\Carleton}{Department of Physics, Carleton University, Ottawa, ON K1S 5B6, Canada}
\newcommand{\CNBCh}{Biological and Chemical Research Centre, Faculty of Chemistry, University of Warsaw, \.Zwirki i Wigury 101, 02-089 Warsaw, Poland}
\newcommand{\Edinburgh}{School of Physics and Astronomy, University of Edinburgh, Edinburgh EH8 9YL, UK}
\newcommand{\Princeton}{Physics Department, Princeton University, Princeton, NJ 08544, USA}
\newcommand{\RHUL}{Department of Physics, Royal Holloway University of London, Egham TW20 0EX, UK}
\newcommand{\TNFBK}{Fondazione Bruno Kessler, Povo 38123, Italy}
\newcommand{\TNTIFPA}{Trento Institute for Fundamental Physics and Applications, Povo 38123, Italy}
\newcommand{\TUM}{Physik Department, Technische Universit\"at M\"unchen, Munich 80333, Germany}
\newcommand{\UCLA}{Physics and Astronomy Department, University of California, Los Angeles, CA 90095, USA}

\author{
  M.\,G.~Boulay\thanksref{LNGS, Carleton}
  \and V.~Camillo\thanksref{LNGS}
  \and N.~Canci\thanksref{LNGS}
  \and S.~Choudhary\thanksref{AstroCeNT}
  \and L.~Consiglio\thanksref{LNGS}
  \and A.~Flammini\thanksref{BOINFN}
  \and C.~Galbiati\thanksref{GSSI, Princeton} 
  \and C.~Ghiano\thanksref{LNGS}
  \and A.~Gola\thanksref{FBK, TIFPA}
  \and S.~Horikawa\thanksref{GSSI} 
  \and P.~Kachru\thanksref{GSSI} 
  \and I.~Kochanek\thanksref{LNGS} 
  \and K.~Kondo\thanksref{LNGS}
  \and G.~Korga\thanksref{RHUL, LNGS}
  \and M.~Ku\'zniak\thanksref{AstroCeNT}
  \and M.~Ku\'zwa\thanksref{AstroCeNT}
  \and A.~Leonhardt\thanksref{TUM}
  \and T.~\L ęcki\thanksref{CNBCh}
  \and A.~Mazzi\thanksref{FBK, TIFPA}
  \and A.~Moharana\thanksref{GSSI}
  \and G.~Nieradka\thanksref{AstroCeNT}
  \and G.~Paternoster\thanksref{FBK, TIFPA}
  \and T.\,R.~Pollmann\thanksref{TUM,currentNikhef}
  \and A.~Razeto\thanksref{LNGS}
  \and D.~Sablone\thanksref{LNGS}
  \and T.~Sworobowicz\thanksref{AstroCeNT}
  \and A.\,M.~Szelc\thanksref{Edinburgh}
  \and C.~T\"urko\u{g}lu\thanksref{AstroCeNT}
  \and H.~Wang\thanksref{UCLA}
}

\institute{
  \AQLNGS\thanksref{email} \label{LNGS}
  \and \Carleton \label{Carleton}
  \and \AstroCeNT \label{AstroCeNT}
  \and \BOINFN \label{BOINFN}
  \and \AQGSSI \label{GSSI}
  \and \Princeton \label{Princeton}
  \and \TNFBK \label{FBK}
  \and \TNTIFPA \label{TIFPA}
  \and \RHUL \label{RHUL}
  \and \TUM \label{TUM}
  \and \CNBCh \label{CNBCh}
  \and \Edinburgh \label{Edinburgh}
  \and \UCLA \label{UCLA}
}
\thankstext{currentNikhef}{Currently at Nikhef and the University of Amsterdam, Science Park, 1098XG Amsterdam, Netherlands}
\thankstext{email}{Corresponding address: sarlabb7@lngs.infn.it}

\journalname{Eur. Phys. J. C}

\begin{document}
\maketitle
\begin{abstract}
A large number of particle detectors employ liquid argon as their target material owing to its high scintillation yield and its ability to drift ionization charge over large distances. Scintillation light from argon is peaked at \SI{128}{\nano\meter} and a wavelength shifter is required for its efficient detection.
In this work, we directly compare the light yield achieved in two identical liquid argon chambers, one of which is equipped with PolyEthylene Naphthalate (PEN) and the other with TetraPhenyl Butadiene (TPB) wavelength shifter. Both chambers are lined with enhanced specular reflectors and instrumented with SiPMs with a coverage fraction of approximately 1\%, which represents a geometry comparable to the future large scale detectors. We measured the light yield of the PEN chamber to be ~39.4$\pm$0.4(stat)$\pm$1.9(syst)\% of the yield of the TPB chamber.
Using a Monte Carlo simulation this result is used to extract the wavelength shifting efficiency of PEN relative to TPB equal to \SI[parse-numbers=false]{47.2\pm5.7}{\percent}.
This result paves the way for the use of easily available PEN foils as a wavelength shifter, which can substantially simplify the construction of future liquid argon detectors.
\end{abstract}

\section{Introduction}
Liquid argon (LAr) based detectors are used in direct dark matter searches, experiments investigating properties of neutrinos and in other applications. LAr acts as an efficient scintillator, however its emission peak is in the vacuum ultraviolet (VUV) at \SI{128}{\nano\meter}. 
Due to lack of photo-sensors highly efficient at VUV wavelengths, 
the scintillation light is typically wavelength shifted (WLS) to the visible range by dedicated films or coatings applied on the detector walls~\cite{WlsReview}.

Production of vacuum evaporated coatings of Tetra-Phenyl Butadiene (TPB)~\cite{Francini}, which is the most commonly used WLS in LAr detectors, requires high vacuum conditions and a process that is not trivial to scale up to hundreds of square meters of surface area, needed for the upcoming generation of experiments.
PolyEthylene Naphthalate (PEN), a polymeric wavelength shifting film available in large formats, has been recently proposed as a scalable and inexpensive alternative to TPB for large LAr-based detectors~\cite{KuzniakPEN}.
Earlier literature-based estimates as well as measurements of wavelength shifting efficiency (WLSE) of PEN relative to that of TPB have suffered from large systematic uncertainty caused by: (1) temperature and excitation-wavelength dependence of the fluorescence process, different in both materials~\cite{KuzniakPEN}, (2) significant sample-to-sample and grade-to-grade variation~\cite{ProtoDPspread} and (3) geometrical effects~\cite{SotoOton}.

In this work we directly compare the light yield (LY) obtained in two identical LAr detector chambers, using PEN or TPB as WLS. This approach allows to reduce the systematic uncertainty on the WLSE of PEN and provides a robust performance test in conditions matching the intended application. For the first time, the geometry used for such comparison is fully representative of large LAr detectors, particularly the planned veto detector of the DarkSide-20k experiment, where photo-sensors cover only a small fraction of the surface~\cite{ESPPU}, and the rest is lined with wavelength-shifting reflectors.

\section{Setup}
\begin{figure}[tb]
\centering
\includegraphics[width=0.35\textwidth]{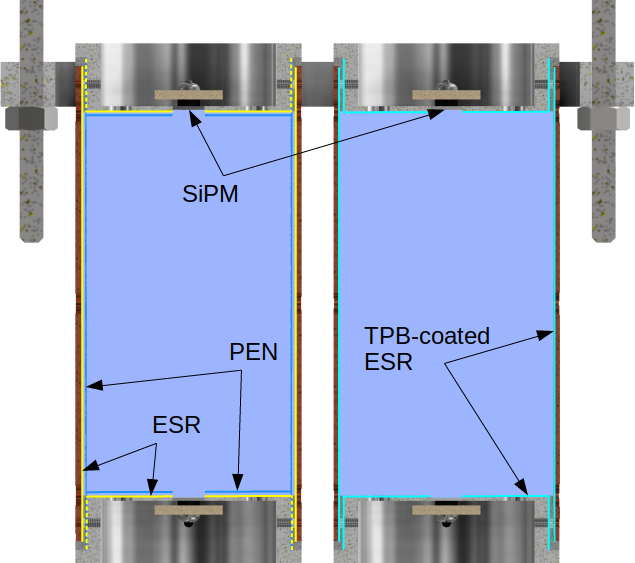}
\caption{Schematic of 2PAC (see text for more details). Sectional view shows the location of photo-detectors and the arrangement of foils: ESR (yellow) and PEN (blue) in the left chamber, TPB-coated ESR (cyan) in the right chamber.}
\label{fig:2pac_schematic}
\end{figure}

A dedicated detector setup has been manufactured for this test, dubbed 2PAC (2 Parallel Argon Chambers), consisting of two cylindrical chambers, with the endcaps instrumented with SiPMs and supplied with spacer rings. The chambers are made out of aluminum and contain a volume of LAr with \SI{47.6}{\milli\meter} diameter and \SI{81}{\milli\meter} length.

The fraction of inner surface covered with photo-sensors ($F$) is approximately \SI{1}{\percent}, while the rest is lined with the Enhanced Specular Reflector (ESR)~\cite{Vikuiti} from 3M, which is a \SI{98}{\percent} reflective polymeric multilayer mirror, commonly used in LAr detectors due to its superior reflectivity to visible light. In one chamber, the ESR is coated with a \SI{3}{\micro\meter} layer of vacuum-evaporated TPB. The other chamber instead contains a \SI{25}{\micro\meter} PEN film (Teonex Q51)\footnote{This grade of PEN has been selected as the most promising candidate in a campaign of ex-situ measurements, to be described in a separate publication.} attached in front of the ESR with up to 1~mm of LAr gap between both films, maintained by spacer rings at the endcaps. Both chambers are fully immersed in a LAr bath and are held with a frame attached to threaded rods suspended from the lid of the cryostat, as shown in Figure~\ref{fig:2pac_schematic}.

The small SiPM coverage forces wavelength shifted photons to be reflected, passing each time through WLS, on average $1/F$ times, before they reach a photo-detector. This makes the overall light yield very sensitive not only to WLSE, but also to effective reflectivity of the ESR/WLS, including the attenuation in WLS itself. 

The motivation behind having a dual chamber system was the possibility to characterize both configurations together in the same LAr bath and with the same radioactive source, thus reducing the systematic uncertainty from varying run conditions.

\subsection{Cryogenics}
\begin{figure}[tb]
\centering
\includegraphics[width=0.45\textwidth]{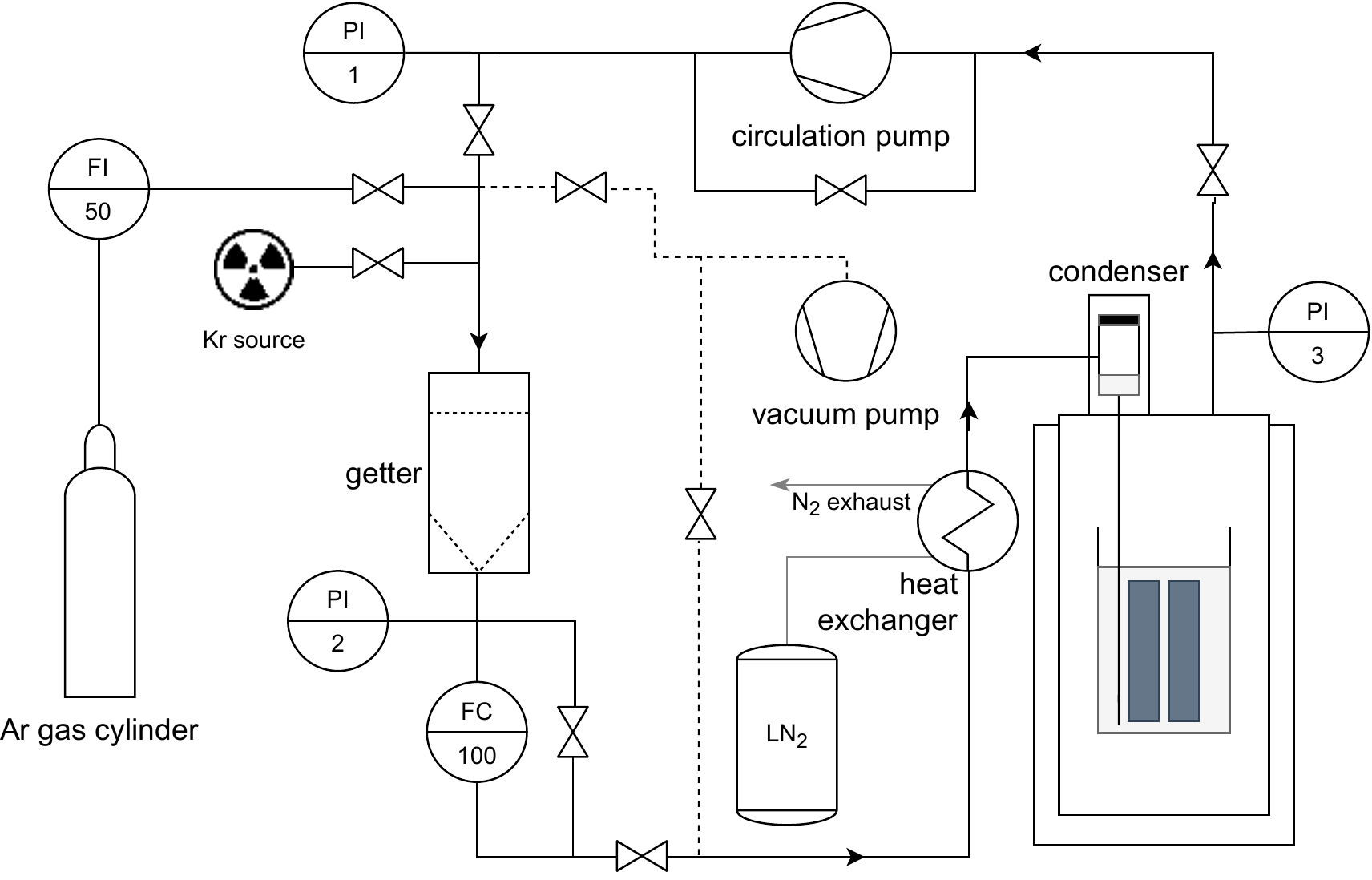}
\caption{The STAR (\underline{S}iPM \underline{T}estbench in \underline{Ar}gon) cryogenic system.}
\label{fig:STARpid}
\end{figure}

The measurement is carried out inside a cryogenic test setup (STAR), which provides a clean liquid argon environment and enables detector tests with the argon scintillation light.
Figure ~\ref{fig:STARpid} shows the whole cryogenic system, which consists of a cryostat and an argon re-circulation circuit.

The cryostat is a vacuum insulated double jacket dewar with an inner diameter of \SI{240}{\milli\meter} and a height of \SI{930}{\milli\meter} and a CF250 top flange from CryoFAB.
It is equipped with an argon gas condenser cooled by a Cryomech PT60 single-stage pulse tube cryo-cooler capable of delivering \SI{60}{\watt} corresponding to a theoretical liquefaction rate of \SI{6}{sl\per\minute}.

A liquid nitrogen powered heat exchanger pre-cools the input argon gas for the condenser from room temperature to about \SI{100}{\kelvin}. The result is an increased liquefaction speed to  \SI{15}{sl\per\minute} with a consumption of about \SI{40}{sl\per\minute} of nitrogen gas.

An inner cylindrical container (\diameter \SI{190}{\milli\meter} by \SI{300}{\milli\meter}) limits the volume of liquid argon necessary to operate the system, significantly reducing the filling time to around \SI{8}{\hour}.
Several PT100 RTDs are installed to monitor the level of liquid argon. 

The argon gas is purified through a SAES PS4-MT3 getter~\cite{saes}, which eliminates impurity in argon gas to sub-ppb level. The argon gas flow is promoted by a circulation pump (Metal Bellows MB-111) and regulated by a Sierra SmartTrack 100 gas flow controller.
The nominal flow in steady state is \SI{5}{sl\per\minute} and the operational pressure of Ar inside the cryostat is about \SI{+50}{\milli\barg}.

The system allows the injection of isotopes of \isotope[83m]Kr into the re-circulation line, as a calibration source with an activity of few tens of becquerels. Such a source is used in several dark matter detectors \cite{ds50first} since it is not filtered by the getter, has a half life of $1.83$ hours (after which it becomes inert) and provides a q-value of \SI{41.5}{\kilo\eV}.

\subsection{Readout}
\begin{figure}[t]
\centering
\includegraphics[trim=0 30mm 50mm 30mm, clip, width=0.23\textwidth]{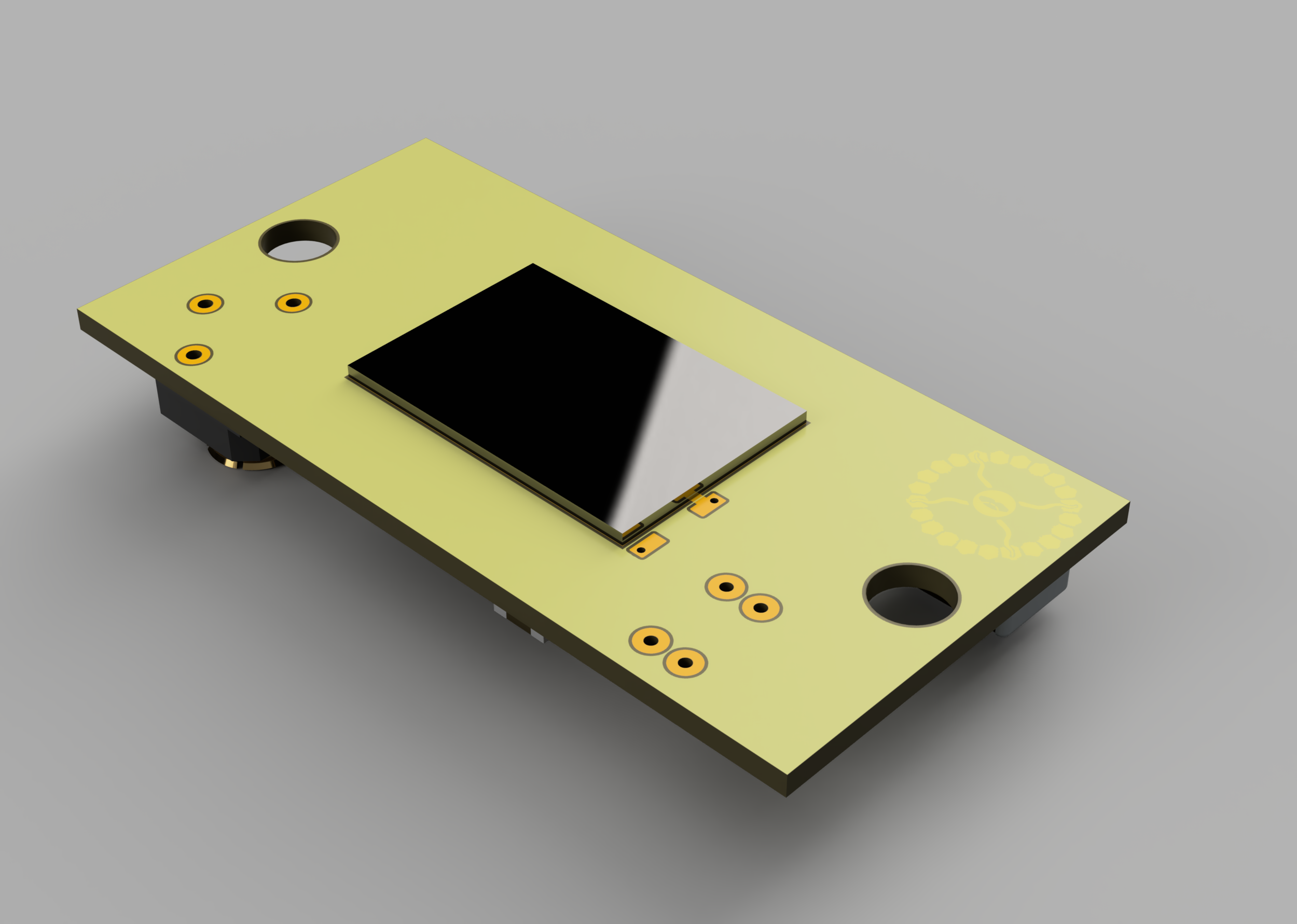}
\includegraphics[trim=40mm 30mm 10mm 30mm, clip, width=0.23\textwidth]{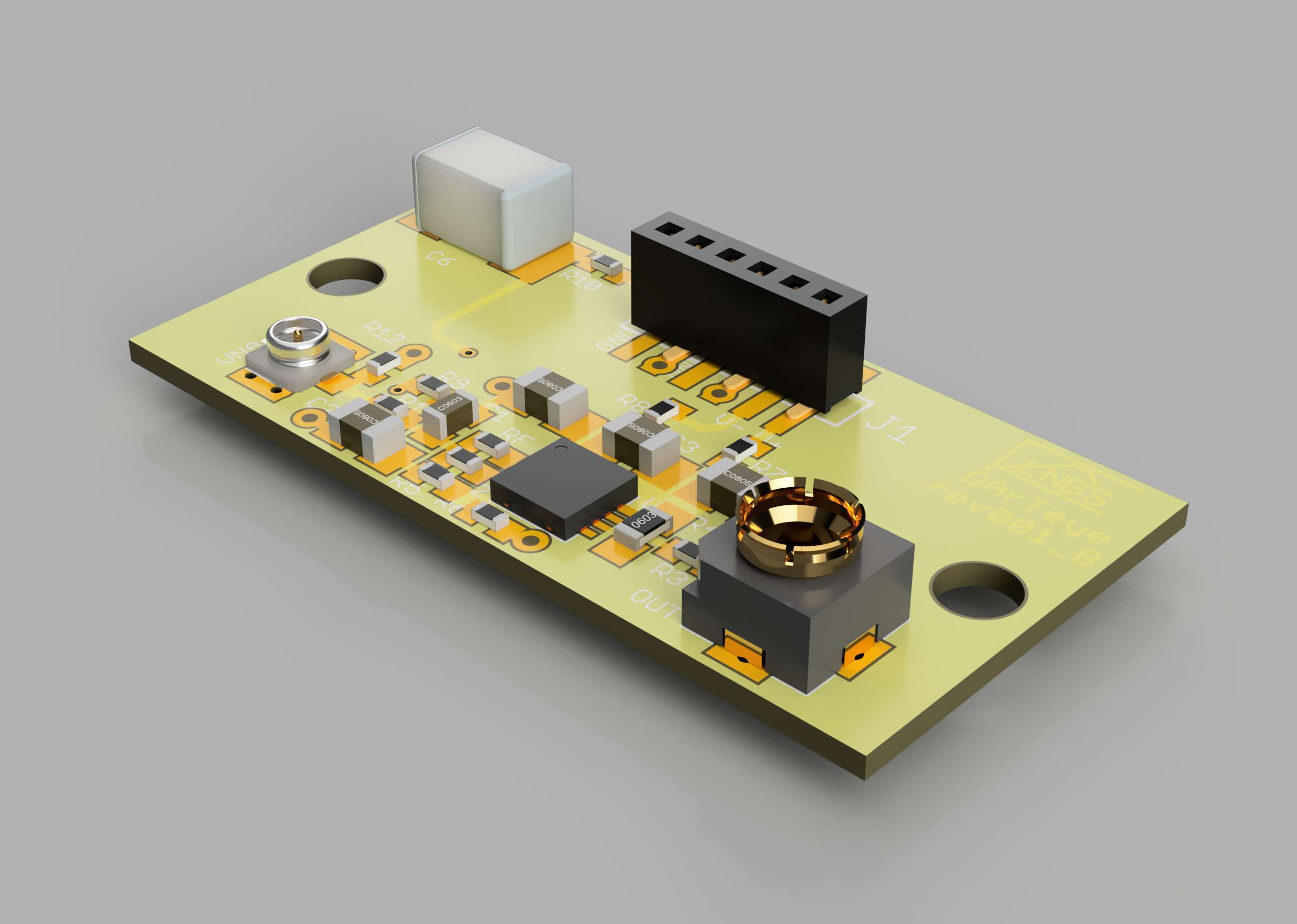}
\caption{The photo-detectors lodge a \SI{1}{\square\cm} SiPM with a low noise cryogenic amplifier in a single \SI[product-units=power]{15 x 30}{\square\milli\meter} board.}
\label{fig:DartEye}
\end{figure}

Photons are detected by cryogenic photo-detectors, designed to incorporate a \SI{1}{\square\cm} SiPM and a low noise pre-amplifier in a compact board, Figure~\ref{fig:DartEye}.  
The units implement the circuit discussed in \cite{cryo-pre}, with a gain of \SI{10.7}{\kilo\volt\per\ampere} (halved by the \SI{50}{\ohm} back termination), an output voltage noise of \SI{350}{\micro\volt} (RMS-AC) for a bandwidth of \SI{36}{\mega\hertz} (at \SI{-3}{\decibel}).
The circuit board is realized in Arlon~55NT to match the coefficient of thermal expansion of silicon~\cite{iza-bari}.

The signals are acquired by a CAEN V1720 unit capable of digitizing \num{8} channels at \SI{250}{\mega S\per\second} with \SI{12}{bits}. A discrete trigger system, based on NIM modules, implements a requirement for a majority of two channels observing at least ten photo-electrons.

The photo-detectors were individually qualified in liquid nitrogen before the installation on the setup.

\subsection{SiPM performances}

The photo-detectors are based on SiPMs of the FBK NUV-HD-Cryo family \SI[product-units=power]{7.9 x 11.7}{\mm} with a single photon avalanche diode cell size of \SI{30}{\micro\meter} and a quenching resistor of \SI{5}{\mega\ohm} (at \SI{86}{\kelvin}). 

We studied the gain of the SiPM (defined as the integral of the generated current) in the range \SIrange{29}{38}{\volt}.
The gain scales linearly with a cell capacitance of \SI{65\pm2}{\femto\farad}, well in agreement with the expectations from the RLC bridge measurements. The breakdown, obtained from the intercept of the gain versus bias, is 
$V_{bd}^C$=\SI{26.8\pm0.1}{\volt}.
The peak amplitude of the amplified signal corresponds to \SI{1.5}{\milli\volt} per volt of bias above the break-down, $V_{bd}^A$=\SI{27.5\pm0.1}{\volt}, at \SI{77}{\kelvin}.

\begin{figure}[tb]
\centering
\includegraphics[width=0.45\textwidth]{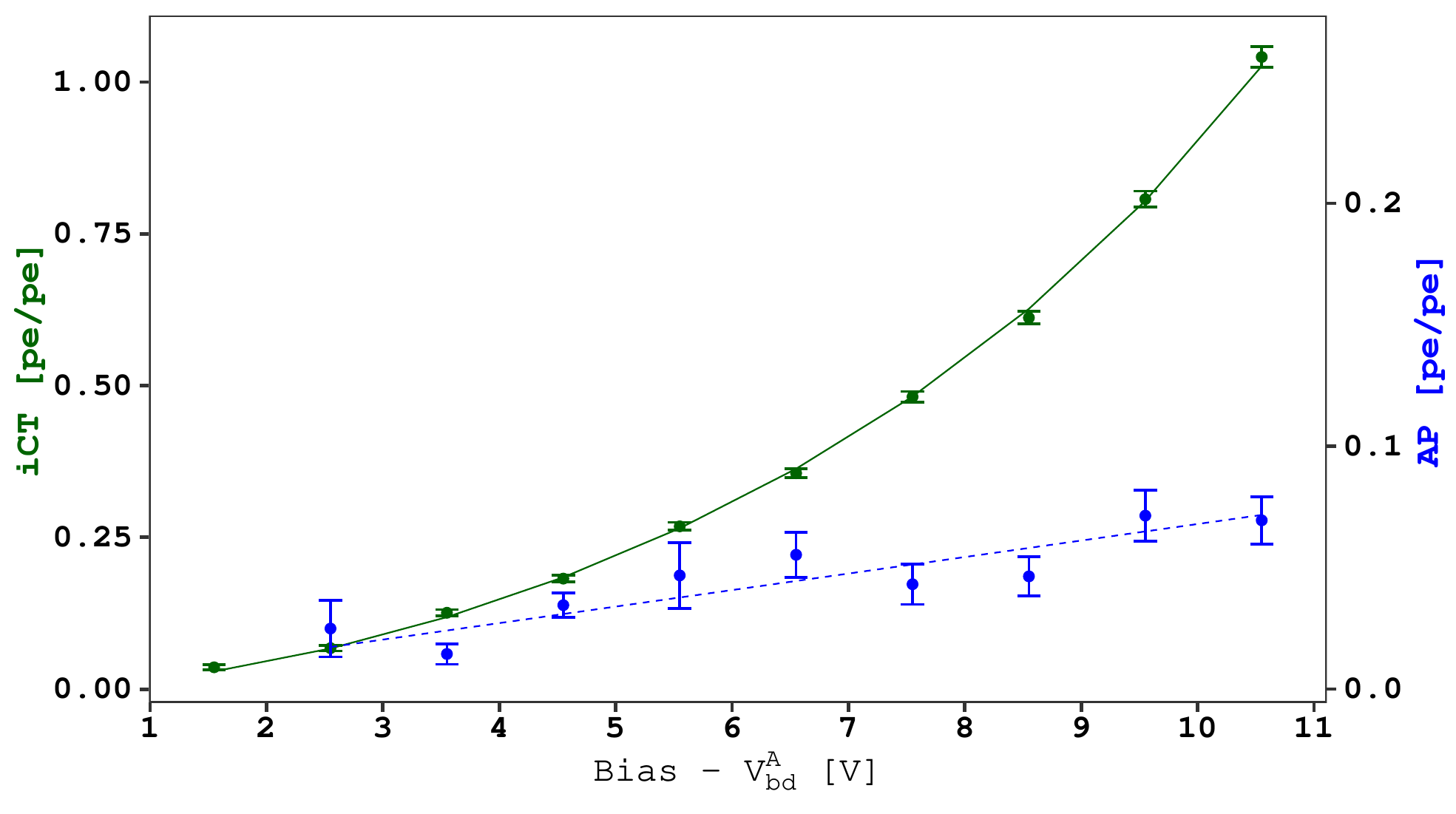}
\caption{
Correlated noises for the NUV-HD-Cryo SiPM at \SI{77}{\kelvin} as excess of photo-electrons measured in a gate of \SI{7}{\micro\second} for single photo-electron events. Lines are drawn to guide the eyes. 
}
\label{fig:correlated_noises}
\end{figure}

SiPM-based photo-detectors are affected by the presence of detection noise: the dark rate (DCR), the cross-talk (iCT) and the after-pulse (AP) \cite{cryo-nuv}. All these elements may potentially affect the signal shape and the reconstructed number of photo-electrons. It is therefore important to quantify them. 
Measurements were performed in a dedicated setup with the same photo-detectors operating in liquid nitrogen to avoid the intrinsic scintillation of argon. Relevant properties of the SiPMs remain stable over the temperature variation between liquid argon and liquid nitrogen~\cite{cryo-nuv, cryo-rgb}.

\label{sec:noises}

The DCR drastically drops below \SI{100}{\kelvin}: we observe a rate of \SI{1}{\cps} per SiPM at the maximum over-voltage. Therefore the contribution to argon scintillation (that lasts a few microseconds) can be safely neglected.
With such a low DCR rate, acquiring enough statistics for the iCT analysis in dark is impractical: therefore we used a short laser pulse to illuminate the SiPM. 
The iCT is calculated by scaling per each over-voltage the average number of acquired photo-electrons by the expected laser occupancy, measured on the empty events. Our data are approximated by the geometric chain cross-talk introduced by Vinogradov~\cite{vinogradov-analytical} that underestimates the iCT at higher over-voltage by a maximum of \SI{10}{\percent}.
The AP is calculated by measuring the relative charge of single photo-electron events in a gate of \SI{7}{\micro\second} after the primary pulse. This definition matches the analysis of the scintillation events which integrates the normalized waveforms. 
The results of these measurements are shown in Figure~\ref{fig:correlated_noises} for both iCT and AP in the over-voltage range of \SIrange{2.5}{10.5}{\volt}.

\section{Analysis}
\begin{figure}[tb]
\centering
\includegraphics[width=0.48\textwidth]{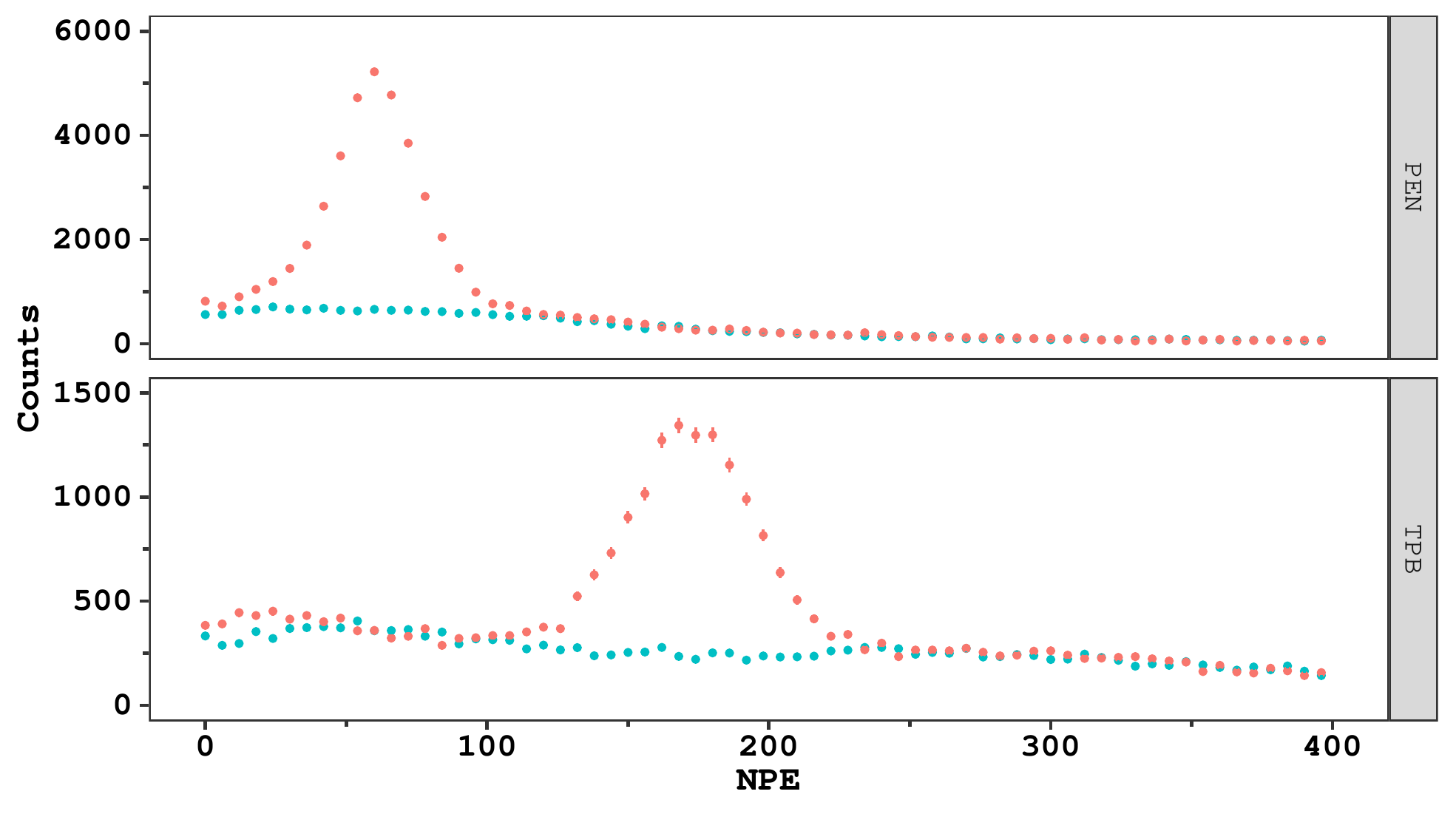}
\caption{\isotope[241]{Am} data (red) and environment data (blue) in the detectors with TPB and PEN, acquired at an over-voltage of \SI{6.5}{\volt}.}
\label{fig:peak}
\end{figure}

An \isotope[241]Am gamma-ray source was used to 
perform the energy calibration of the system. As a cross-check in Run 2 an additional calibration with a \isotope[83m]Kr source was performed. Calibrating the detectors with two sources allowed us to confirm the linearity of the energy scale. The \isotope[241]Am source was fixed on the outer surface of the Dewar, at a location corresponding to the center of the detector system. The \SI{59.5}{\kilo\eV} gammas penetrate the liquid argon buffer outside the 2PAC detectors (reaching the active volumes) and exhibit a high cross-section for the full absorption peak. The \isotope[83m]Kr is diffused uniformly inside the detector~\cite{Kr-mixing-paper} and decays with two transitions with a half-life of \SI{154}{\nano\second}.

In the analysis, we integrate the normalized waveforms for a gate of \SI{7}{\micro\second}. 
In such a window \SI{99}{\percent} of the scintillation light is detected (including the ballistic deficit due to the finite recharge time of the SiPMs). Absorption and re-emission by the WLS and delays introduced by the optical paths inside the detectors have no effect on this timescale. 
The normalization procedure includes removal of the baseline, calculated in the pre-trigger, and scaling by the gain of the photo-detectors. The SiPM correlated noises are accounted for statistically by dividing the integrals by the factor $(1+\textrm{AP})(1+\textrm{iCT})$, where the values of AP and the iCT are indicated in Figure~\ref{fig:correlated_noises}.

Data were acquired without the \isotope[241]Am and \isotope[83m]Kr sources to subtract the environmental background and isolate better the full absorption peaks, which are then fitted with a Gaussian model, as shown in Figure~\ref{fig:peak}. From the fit we extract the mean number of photo-electrons for the sum of both photo-detectors in each chamber. The light yield is given by the ratio of the detected photo-electrons and the energy deposited in the medium. The light yield values obtained with both sources were found to be consistent with each other.

\begin{figure}[tb]
\centering
\includegraphics[width=0.48\textwidth]{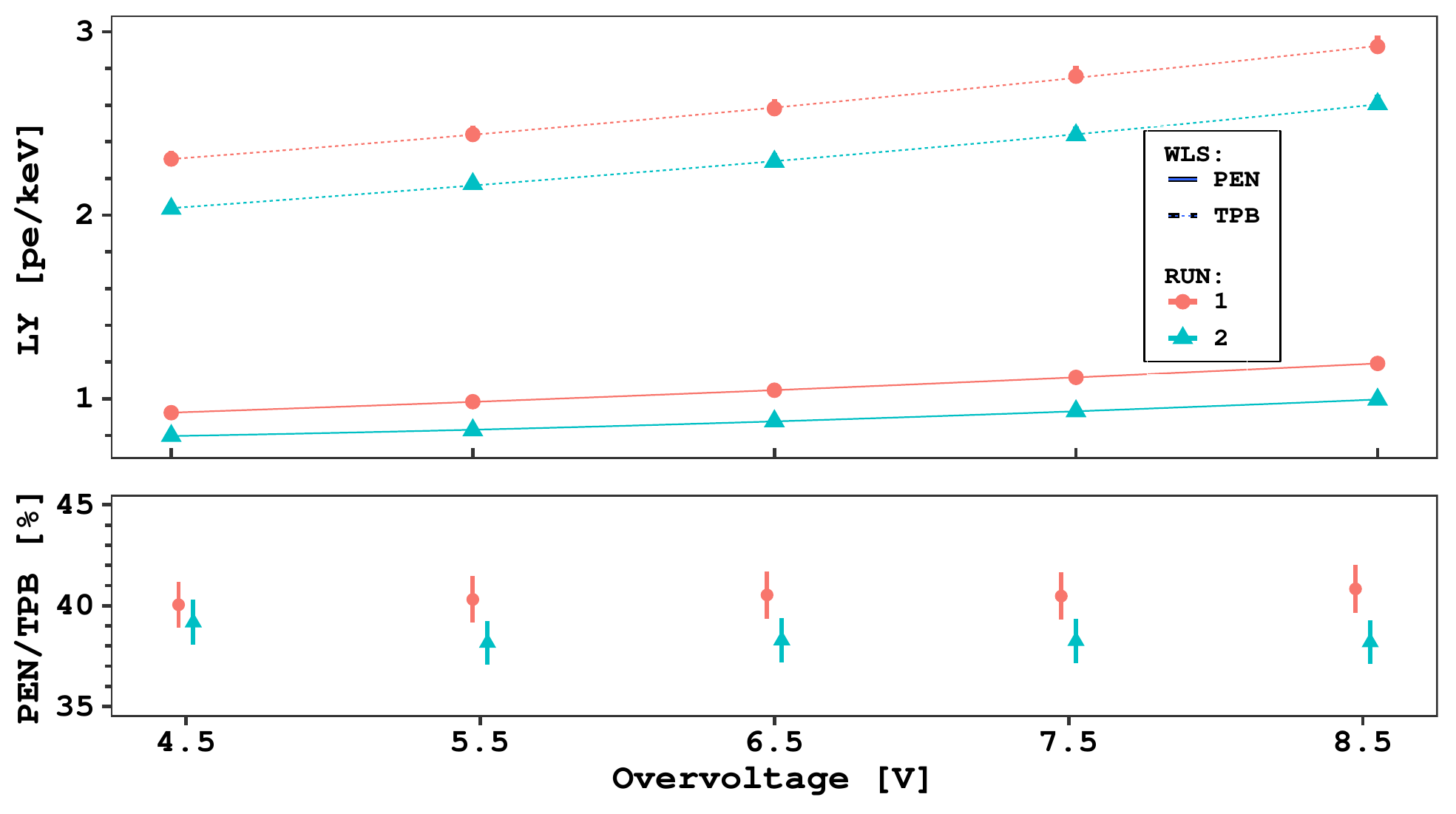}
\caption{Light yield of the chambers equipped with PEN and TPB (top panel) and their ratio (bottom panel) for two experiments runs. }
\label{fig:ly}
\end{figure}

One of the photo-detectors (bottom of TPB chamber) suffered from reduced light collection efficiency due to window misalignment. The size of the effect was evaluated with a Monte Carlo (see later) based on as-built geometry and factored into the reported light yield as a correction for the raw charge spectrum of the interested photo-detector. This correction increases the LY of the TPB chamber by \SI{6}{\percent}.

To check the PDE consistency of all four SiPMs (which were extracted from the same silicon wafer) a second run was acquired with top photo-detectors of the two chambers swapped. The top/total asymmetry was at the level of \SI{50\pm1}{\percent} for both chambers and in both runs, proving the homogeneity of the SiPMs (and the goodness of the correction for the misaligned window).

The results are shown in Figure~\ref{fig:ly}: for the second run the light yield is lower by about \SI{15}{\percent}. This can be explained by the different purity of the argon in the system, e.g. a sub-ppm level of Nitrogen contamination~\cite{Acciarri_2010}. Monitoring of the LAr purity and triplet lifetime sufficiently sensitive to confirm such level of contamination exceeds the current capabilities of the setup.

\section{Simulation}
\label{sec:mc}
As wavelength shifted photons reflect from the walls many times before reaching photosensors, the final LY strongly depends not only on WLSE, but also on the reflectivity of all inner surfaces. In order to decouple both effects and evaluate the PEN WLSE, a Monte Carlo model of 2PAC was implemented in Geant4, taking into account its geometry and optical properties of all relevant materials: liquid argon, ESR, TPB, PEN and SiPMs, as well as the geometry and materials of the surrounding cryostat. The Monte Carlo results were validated against a simplified analytic model for estimation of the light yield of scintillation detectors~\cite{Segreto_2012}.
\begin{figure}[tb]
\centering
\includegraphics[width=0.45\textwidth]{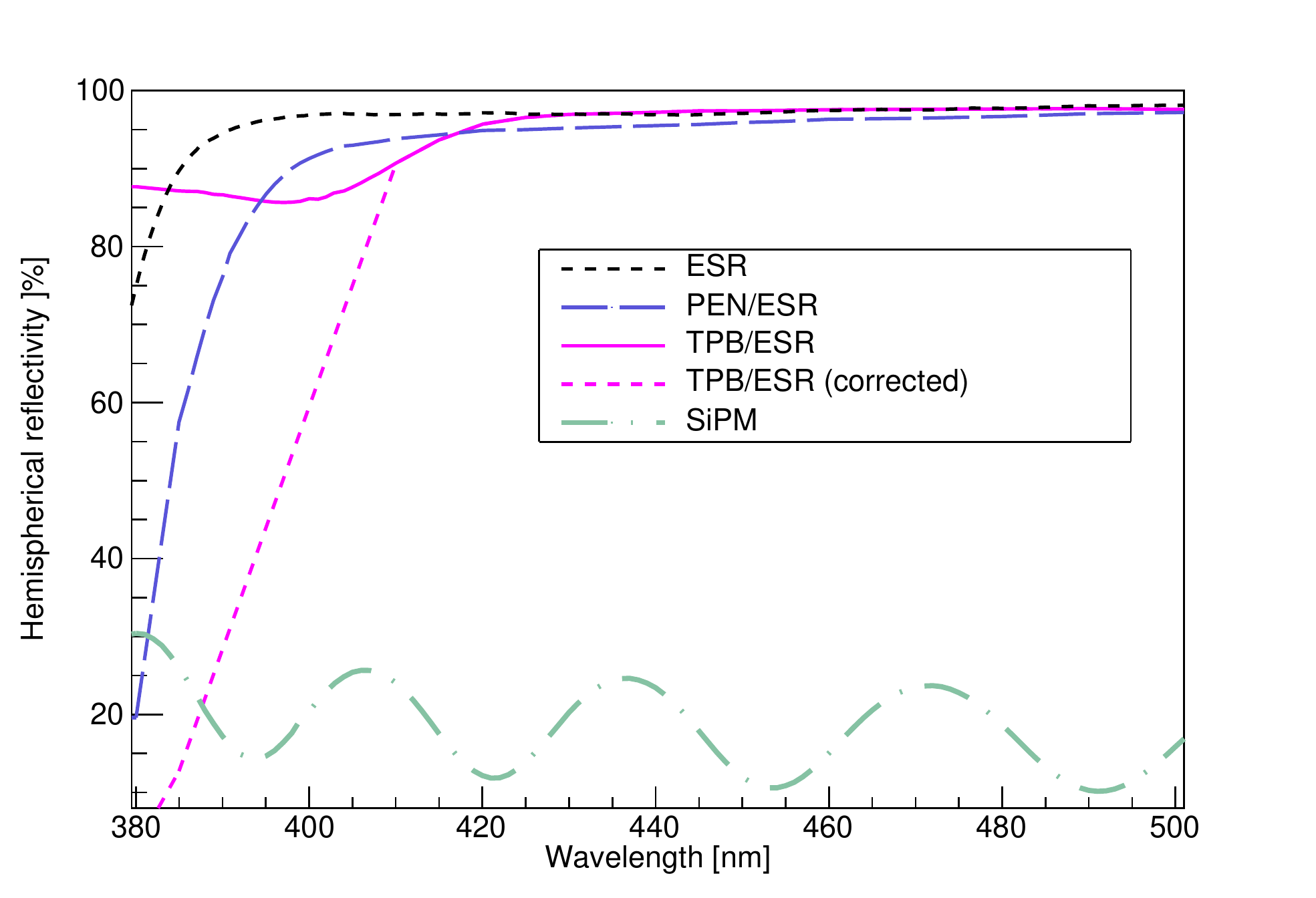}
\caption{Hemispherical reflectivity measured at 7$^\circ$ angle of incidence with a spectrophotometer equipped with an integrating sphere for: ESR, PEN air-coupled to ESR, TPB evaporated on ESR, TPB evaporated on ESR corrected for the spurious fluorescence component based on~\cite{Francini}, and SiPMs (see legend). 
}
\label{fig:reflectivity}
\end{figure}

The reflectivity spectra used in the simulation for ESR, TPB-coated ESR, PEN air-coupled to ESR and SiPMs have been measured using a Shimadzu UV-3600 spectrophotometer equipped with a 15~cm diameter integrating sphere accessory (LISR-3100), collecting the light reflected into the entire hemisphere, see Fig.~\ref{fig:reflectivity}. The measurements were performed relative to a BaSO$_4$ reference, with \SI{98.0\pm0.5}{\percent} absolute reflectivity in the visible range.

Literature values for other parameters have been used, namely the emission spectra of PEN~\cite{Mary:1997} and TPB~\cite{Francini} at 87~K. Absorption of VUV light in LAr and of visible light in LAr and in TPB was considered negligible.

A range of values for the TPB WLSE is present in the literature~\cite{WlsReview}: in the simulation it was fixed at \SI{100}{\percent}, since the relative performance of both WLS was of main interest. In the relative measurement presented here, factors identically affecting the absolute LY for both configurations cancel out, particularly the LAr scintillation yield and the SiPM PDE scaling. 

The PDE for SiPMs, taken at \SI{6}{\volt} of over-voltage, was measured at room temperature by FBK, similar to the published result~\cite{ds20k}, and its values in LAr are not known. 
For the parameters described above, the simulation predicts a light yield of \SI{2.6\pm0.5}{pe\per\kilo\eV} for the TPB chamber, consistent with the corresponding measurement, interpolated to \SI{2.5\pm0.05}{pe\per\kilo\eV}. 
The scaling of the PDE cancels out in the determination of the relative PEN WLSE (as discussed above), however the change of shape of the PDE curve with respect to the room temperature data could bias the result. In the Monte Carlo we simulated a shift of PDE by \SI{20}{\nano\meter} towards lower wavelengths as a result of the increased extinction coefficient of visible light in silicon at cryogenic temperature~\cite{dash-newman}. This shift produces an effect of about \SI{2}{\percent} that is negligible in respect to the other uncertainties.
\begin{table}[b]
\footnotesize
\centering
\begin{tabular}{l|c|c}
{\bf Input} & {\bf Allowed range} & \boldmath$\delta_{syst}/\frac{LY_{PEN}}{LY_{TPB}}$ \\
\hline
$R$ calibration & 0.980$\pm$0.005 & 0.07 \\
(reference) & & \\
\hline
$R$ correction & 0.994$\pm$0.006 & 0.08 \\
(slits and bolts) & & \\
\hline
$R$ in LAr &  1.000$\pm$0.0035 & 0.05 \\
\hline
TPB $R$ spectrum & extrapolate & 0.02 \\
(spurious component & below 415~nm& \\
removal) & &\\
\hline
SiPM reflectivity & 0.171$\pm$0.017 (PEN) & 0.01 \\
& 0.167$\pm$0.017 (TPB) & 0.01 \\
\hline
PDE curve shape & shift by --20 nm  & 0.02 \\
$T$ dependence & &\\ 
\hline
{\bf Total} & & {\bf 0.12}
\end{tabular}
\caption{Systematic uncertainties on key inputs to the Monte Carlo simulation and their relative effect on the final LY ratio of PEN and TPB configurations. See text for more details.}
\label{tab:systematic}
\end{table}

Table~\ref{tab:systematic} lists the systematic uncertainties on input parameters for the Monte Carlo and their effect on the simulated LY ratio of PEN and TPB configurations. The dominant factors are the uncertainty on the absolute calibration of the reflectivity measurements, and on the losses due the fraction of 2PAC surface not covered with the ESR film.

The SiPM reflectivity is non-trivial to model due to the anti-reflective coating. 
We used our own measurement, shown in Fig.~\ref{fig:reflectivity} and estimated the maximum variation of reflectivity between air and LAr for bare silicon.
The final curve used in the model was the mean of the air and LAr-corrected values, with a systematic error assigned to cover both extremities.

To evaluate the reflectivity change of PEN/ESR, TPB/ESR and ESR in LAr with respect to ex-situ measurement in air, the optical stack model of ESR used in Ref.~\cite{LOIGNONHOULE201762} was extended to accommodate an additional film layer, a separating gap, and the LAr medium. The effect due to the change of the medium averaged over the respective WLS emission spectrum is small and was included as a systematic.

In the integrating sphere measurement, the reflectivity curve of TPB-coated ESR below \SI{415}{\nano\meter} is affected by a spurious wavelength shifted component, 
therefore we extrapolated our data as described in ~\cite[Figure~18 therein]{Francini}. Similar effect for PEN takes place only below 380~nm where the overall reflectivity is already low and has therefore negligible impact on the result.

PEN bulk absorption length for Teonex Q51 films has not been measured directly, however the integrating sphere reflectivity measurement of PEN/ESR used in the simulation (blue dashed line in Fig.~\ref{fig:reflectivity}) combines reflection, refraction and absorption effects. To avoid accounting for them twice, in the simulations the refractive index of PEN was set equal to that of LAr and the absorption in PEN was disabled. As a cross-check, a separate simulation was performed following the microscopic rather than the effective approach, where instead the literature refractive index and absorption length of PEN were used, together with the reflectivity of bare ESR (black dotted line in Fig.~\ref{fig:reflectivity}).
The attenuation length varies about 1~cm near the peak region (425~nm to 475~nm), as reported for injection moulded PEN tiles~\cite{Efremenko_2019}; for the cross-check a constant 1~cm value was set in the simulation. Both results agree to 1\%, and varying the absorption length by 1~cm changes the result approximately by the size of the overall systematic uncertainty. The Monte Carlo results are robust against sub-mm non-uniformities in the size of the LAr gap between PEN and ESR.

Finally, we consider the recently reported effect of TPB `emanation' from evaporatively coated ESR foils, potentially biasing the relative measurement through presence of non-zero concentration of TPB dissolved or suspended in LAr inside of the PEN chamber. In Ref.~\cite{Asaadi_2019} 2~ppb molar fraction of TPB in LAr was observed after 24-72~h soak of a 103~cm$^2$ foil in 119~cc of LAr. In 2PAC VUV photons travel on average 2~cm prior to reaching PEN. Assuming conservatively that similar concentration is reached in 2PAC (157~cm$^2$ of the foil in 8.5~l of LAr) within 40-140~h of data taking and applying the TPB extinction coefficient at 200~nm of 2$\times$10$^4$~(M$\cdot$cm)$^{-1}$~\cite{WallaceWilliams_1994}, one arrives at $<$0.6\% contribution from VUV light converted in TPB. Therefore, the effect is small in comparison with other systematic uncertainties and neglected in the analysis.

In order to simulate LY for both configurations, monoenergetic 59.5~keV gammas were generated from a point on the outer cryostat wall, corresponding to the actual geometry of the experimental setup. In the simulation gammas are propagated through the materials of the cryostat, and then scintillation photons induced by energy deposits in LAr are tracked, taking into account wavelength shifting, attenuation, reflection and detection processes. A Gaussian fit to the full absorption peak in the photoelectron spectrum was used to evaluate the simulated LY, similarly as in the analysis of the detector data.

\section{Results}
Figure~\ref{fig:ly} reports the ratio of the light yields of the two chambers. Weighted average of such ratio, for all over-voltages and runs, results \SI[parse-numbers=false]{39.4\pm0.4\textrm{(stat)}\pm1.9\textrm{(sys)}}{\percent}, where the systematic uncertainty includes contributions from the residual top/bottom asymmetry and the difference between the runs. 
The ratio is affected by the reduced LY of the second run by less than \SI{2}{\percent}: the quoted value is the average ratio from both runs.

We used the MC simulation, which accounts for the geometry and optics of the system, to evaluate the relative WLSE of PEN with respect to TPB from the observed LY. The systematic uncertainties on the input parameters have been studied and summarized in Table~\ref{tab:systematic}. Systematic uncertainty on the result from all contributions have been added in quadrature and propagated into the final result, derived as 
\begin{equation}
\frac{WLSE_{PEN}}{WLSE_{TPB}} = \left(\frac{LY^{exp.}_{PEN}}{LY^{exp.}_{TPB}}\right)/\left(\frac{LY^{sim.}_{PEN}}{LY^{sim.}_{TPB}}\right).
\label{eq-wlse}
\end{equation}
The resulting value for WLSE of PEN relative to TPB for the \SI[parse-numbers=false]{25}{\micro\meter} thick Teonex Q51 foil is 47.2$\pm$5.7\%, where the uncertainty is dominated by the systematic error. This is a significantly higher result than the previously projected/measured results for other PEN grades, particularly \SI[parse-numbers=false]{34.0\pm1.1}{\percent} for \SI[parse-numbers=false]{125}{\micro\meter} thick Teonex~Q53 \cite{Abraham:2021otn} or \SI[parse-numbers=false]{38.0\pm7}{\percent} for \SI[parse-numbers=false]{125}{\micro\meter} thick~Teonex Q83~\cite{KuzniakPEN}.

\section{Conclusions}
Scalability and simplicity of PEN motivates its use as an alternative WLS in large LAr detectors. Benchmarking its performance against commonly used TPB is challenging due to temperature and excitation wavelength dependence as well as non-trivial geometrical effects. The 2PAC detector, built to emulate in small scale the key features of a large LAr detector (walls lined with WLS reflector and SiPM coverage fraction of approximately 1\%) allowed, for the first time, for a meaningful comparison of both materials in conditions equivalent to the end-goal application: in LAr and with the 4$\pi$ light collection.

The ratio of LYs obtained in 2PAC chambers is a robust result which can be used to model the yield of larger LAr detectors equipped with SiPMs and with reflector and SiPM coverage fractions comparable with 2PAC. Employing a Monte Carlo simulation we extracted the value of PEN WLSE relative to TPB, which exceeded that of those recently reported in the literature for other PEN grades.

We note that Teonex~Q51 has no planarizer coating (contrary to Q65~\cite{MacDonald2007}), and does not undergo high temperature annealing step during production (which is the case for Q81, Q83 and Q65 grades, in order to reduce their shrinkage properties), which may affect the film crystallinity. Finally, lower film thickness, in comparison to~\SI[parse-numbers=false]{125}{\micro\meter} used previously, reduces losses from self-absorption. Detailed results of a survey of available PEN grades will be published separately.

The PEN WLSE result is sufficient for many applications and can now be reliably used to simulate arbitrary detector configurations. Given that such a result was obtained with a technical grade of PEN, which was industrially mass-produced and not optimized for wavelength shifting of LAr scintillation light, it is compelling to further develop PEN and similar polymeric materials for applications as WLS in the future very large detectors, such as ARGO or DUNE, where WLS surfaces between hundreds and thousands of square meters will be needed.

\section*{Acknowledgements}
This work was supported from the European Union’s Horizon 2020 research and innovation programme under grant agreement No 952480 (DarkWave project).
The 2PAC detector construction was carried out with the use of CEZAMAT (Warsaw) cleanroom infrastructures financed by the European Regional Development Fund; we are grateful to Maciej Trzaskowski of CEZAMAT for support.
We are grateful to Prof. Magdalena Skompska for access to the spectrophotometer, which was purchased by CNBCh (University of Warsaw) from the project co-financed by EU from the European Regional Development Fund.
We acknowledge support from the Istituto Nazionale di Fisica Nucleare (Italy) and Laboratori Nazionali del Gran Sasso (Italy) of INFN, from NSF (US, Grant PHY-1314507 for Princeton University), from the Royal Society UK and the Science and Technology Facilities Council (STFC), part of the United Kingdom Research and Innovation, and from the International Research Agenda Programme AstroCeNT (MAB\allowbreak/2018\allowbreak/7) funded by the Foundation for Polish Science (FNP) from the European Regional Development Fund.
Fruitful discussions with Gemma Testera are gratefully acknowledged.

\bibliographystyle{spphys}
\bibliography{main}

\end{document}